%% file: main.tex
\documentclass{article}

\usepackage{PRIMEarxiv}

\usepackage[utf8]{inputenc} 
\usepackage[T1]{fontenc}    
\usepackage{hyperref}       
\usepackage{url}            
\usepackage{booktabs}       
\usepackage{amsfonts}       
\usepackage{nicefrac}       
\usepackage{microtype}      
\usepackage{lipsum}
\usepackage{fancyhdr}       
\usepackage{graphicx}       
\graphicspath{{media/}}     

\usepackage{times}  
\usepackage{titlesec}
\usepackage{tikz}
\usepackage{amsmath}
\usepackage{xspace}
\usepackage{xcolor}
\usepackage[T1]{fontenc}
\usepackage{subfigure}
\usepackage{verbatim}
\usepackage{url}
\usepackage{enumitem}
\usepackage{listings}
\usepackage{multirow}
\usepackage{multicol}
\usepackage{array}
\usepackage{diagbox}
\usepackage[ruled,linesnumbered]{algorithm2e}
\usepackage{pifont}
\usepackage{mathtools}
\usepackage{fancyhdr}
\usepackage{caption}
\usepackage{xurl}

\newcommand{\sysname}{\texttt{MegatronApp}\xspace}
\newcommand{\fbd}{\texttt{MegaFBD}\xspace}
\newcommand{\dpp}{\texttt{MegaDPP}\xspace}
\newcommand{\sd}{\texttt{MegaScan}\xspace}
\newcommand{\vs}{\texttt{MegaScope}\xspace}

\newcommand{\para}[1]{\smallskip\noindent\textbf{#1}}

\title{\sysname: Efficient and Comprehensive Management on Distributed LLM Training
}

\author{
  Bohan Zhao, Yongchao He \\
  Suanzhi Future \\
  \texttt{contact@siflow.cn} \\
   \And
  Guang Yang, Shuo Chen, Ruitao Liu, Tingrui Zhang, Wei Xu \\
  Shanghai Qi Zhi Institute \\
  \texttt{xuwei@sqz.ac.cn} \\
}

\begin{document}
\maketitle

\begin{abstract}
The rapid escalation in the parameter count of large language models (LLMs) has transformed model training from a single-node endeavor into a highly intricate, cross-node activity. While frameworks such as Megatron-LM successfully integrate tensor (TP), pipeline (PP), and data (DP) parallelism to enable trillion-parameter training, they simultaneously expose practitioners to unprecedented systems-level challenges in performance optimization, diagnosis, and interpretability. \sysname is an open-source toolchain expressly designed to meet these challenges. It introduces four orthogonal, yet seamlessly composable modules--MegaScan, MegaFBD, MegaDPP, and MegaScope--that collectively elevate the reliability, efficiency, and transparency of production-scale training. This paper presents the motivation, architecture, and distinctive contributions of each module, and elucidates how their synergistic integration augments the Megatron-LM ecosystem.
\end{abstract}


\input{intro}

\input{background}

\input{MegaScan}

\input{MegaFBD}

\input{MegaDPP}

\input{MegaScope}

\input{Evaluation}

\input{Conclusion}

\bibliographystyle{unsrt}  
\bibliography{references}

\end{document}

%% file: intro.tex
\section{Introduction}
Large-scale transformer~\cite{vaswani2017transformer} training now operates at cluster scale, where thousands of accelerators coordinate over bandwidth-constrained networks and failure-prone hardware. Minor perturbations such as transient GPU throttling or link jitter can cascade into significant slowdowns, while the opaqueness of distributed execution hampers root-cause analysis. At the same time, the interpretability community increasingly demands fine-grained introspection~\cite{shi2025routesparseautoencoderinterpret,marks2024interpretinglearnedfeedbackpatterns,cunningham2023sparseautoencodershighlyinterpretable,lindner2023tracrcompiledtransformerslaboratory,zimmermann2024measuringperunit} of model states during training--yet naïvely capturing activations or attention maps in trillion-parameter models~\cite{liu2024deepseek,dubey2024llama3.1, yang2025qwen3} quickly saturates I/O subsystems.

Existing profiling, monitoring, and visualization tools~~\cite{vig2019bertviz,TensorBoard,chen2020mlflow} provide partial relief but suffer three key limitations: (i) lack of temporal causality modeling, rendering them ineffective at distinguishing sources from victims of performance anomalies; (ii) insufficient awareness of deep-learning communication semantics, resulting in high false-positive rates under structured parallelism~\cite{narayanan2021megatron,korthikanti2023megatron2}; and (iii) rigid data-collection pipelines that either couple tightly to user code or impose prohibitive overheads.
\sysname addresses these deficiencies through a suite of lightweight extensions that require only minimal code changes yet expose rich, semantically meaningful telemetry.

We summarize our contributions as follows:

\sd: a CUDA-event-driven tracing engine that performs on-line slow-node detection and root-cause localization at operator granularity.

\fbd: a forward-backward decoupling layer that reallocates the two phases across heterogeneous resources, mitigating memory contention and improving utilization.

\dpp: a dynamic pipeline scheduler that adapts traversal order and communication overlap to network and compute volatility.

\vs: a pluggable visualization and perturbation framework that enables low-overhead capture, compression, and interactive rendering of intermediate tensors.

Each component is optional, self-contained, and activated via simple runtime flags, preserving compatibility with upstream Megatron-LM. \sysname is publicly available at \url{https://github.com/OpenSQZ/MegatronApp}.

%% file: background.tex
\section{Background}

\subsection{The Era of Trillion-Parameter LLMs}

Model size has grown exponentially since the original transformer, vaulting from millions to \emph{trillions} of parameters in under a decade.  NVIDIA's Megatron project~\cite{shoeybi2019megatron1,narayanan2021megatron, korthikanti2023megatron2} first demonstrated that a GPT-style model with one trillion parameters could be trained in practicable wall-clock time by combining tensor, pipeline, and data parallelism across 3072~A100 GPUs.  
This escalation in scale makes distributed training not a luxury but a necessity: even if a single accelerator had enough memory, compute time would be prohibitive (e.g., 36 years on eight V100s for GPT-3~\cite{li2014ps}.

\subsection{Distributed Training Fundamentals}

Modern LLM training relies on three orthogonal parallelism strategies:

\begin{itemize}
    \item \textbf{Data parallelism (DP)}~\cite{li2014ps, patarasuk2009allreduce} replicates the model and splits the batch across devices, synchronizing gradients with \texttt{AllReduce}.  Its conceptual simplicity yields near-linear scaling until memory limits bite.  
    \item \textbf{Tensor parallelism (TP)}~\cite{narayanan2021megatron} partitions weight matrices so that a single layer spans multiple GPUs, reducing per-device memory at the expense of tighter communication coupling.  
    \item \textbf{Pipeline parallelism (PP)}~\cite{huang2019gpipe,narayanan2019pipedream,narayanan20212bw} allocates consecutive layer blocks to different devices and overlaps micro-batch execution; bubble overhead and activation storage dominate its efficiency.  
\end{itemize}

State-of-the-art systems combine all three dimensions--commonly dubbed \emph{3-D parallelism}--to scale well beyond 10\,B parameters.

\subsection{Ecosystem Landscape}

\paragraph{Megatron-LM.}  Originally introduced by NVIDIA, Megatron-LM~\cite{narayanan2021megatron} remains the de-facto reference for GPT-style training with tight integration of TP, PP, and DP.

\paragraph{DeepSpeed and ZeRO.}  Microsoft's DeepSpeed~\cite{DeepSpeed} library tackles optimizer and activation memory via ZeRO-3/Infinity offloading~\cite{ren2021zero} and NVMe staging, enabling 500 B-scale models on modest clusters.

\paragraph{PyTorch FSDP and Others.}  PyTorch 2.3 stabilised Fully-Sharded Data Parallel (FSDP)~\cite{zhao2023fsdp}, bringing zero-redundancy sharding into the core framework, while projects such as MosaicML Composer and Hugging Face Accelerate target ease-of-use for sub-100 B models.

\subsubsection{LLM Interpretability}

A growing body of work seeks to open the "black box'' of Transformer-based LLMs.  Zhao \emph{et al.} offer one of the earliest systematic taxonomies, covering local- and global-explanation techniques, attribution methods, and probing tasks~\cite{Zhao2023ExplainabilitySurvey}.  Luo and Specia extend this line with a 2024 survey that also catalogues practical uses of explanations--e.g., model editing and controllable generation--highlighting the tension between faithfulness and usability~\cite{Luo2024LLMExplainability}.  Mechanistic-interpretability efforts~\cite{Nanda2023MechInterpSurvey, Meng2023ROME, Meng2023MEMIT} (e.g.\ neuron activation patching, causal tracing) complement these surveys but remain nascent for trillion-parameter scales, motivating the interactive visualization that \texttt{MegaScope} supplies.

\subsubsection{Pipeline-Parallel Scheduling}

Pipeline parallelism amortises memory across stages but suffers from bubble overhead and load imbalance.  A 2024 comprehensive review contrasts synchronous (\textsc{GPipe}~\cite{huang2019gpipe}) and asynchronous (\textsc{PipeDream}~\cite{narayanan2019pipedream}) schedules, emphasising stage-wise profiling and recomputation to minimise idle time~\cite{Guan2024PipelineOverview}.  More recently, \textsc{DawnPiper}~\cite{Peng2025DawnPiper} introduces a compilation-based profiler and dynamic chunking to shrink stage memory while keeping utilization high in multi-GPU clusters. Qi \emph{et al.} propose Zero Bubble Pipeline Parallelism (ZBPP) to (almost) eliminate pipeline bubbles by carefully interleaving forward/backward steps and weight-gradient computations~\cite{Qi2024ZBPP}. 
DualPipe introduces a \emph{bidirectional, dual-channel} pipeline that \emph{fully overlaps} forward/backward computation with communication~\cite{liu2024deepseek}.
These studies underscore the need for runtime visibility into per-stage latency--functionality built into \texttt{MegaScope} and exploited by \texttt{MegaDPP} for adaptive scheduling.

\subsubsection{Straggler Detection and Mitigation}

Stragglers--GPUs that throttle or experience link jitter--can slash throughput by cascading delays through tightly coupled collectives.  A 2025 ByteDance trace study quantifies this impact, using what-if analysis to show that removing the slowest 5\% of iterations yields up to 38\% speed-ups~\cite{Lin2025StragglerWhatIf}.  Earlier, \textsc{DPro-SM} proposed LSTM-based runtime prediction to pre-emptively migrate work away from prospective stragglers~\cite{Ravikumar2024DProSM}. 
\textbf{Greyhound} takes a complementary, production-driven angle: it first \emph{characterises} more than 10 000-GPU jobs on an industrial cluster and finds that \emph{fail-slows}--transient, sub-minute-to-multi-hour stragglers--inflate job runtime by 1.34$\times$ on average~\cite{Wu2025Greyhound}.  
These works reinforce the importance of fine-grained latency telemetry, which \texttt{MegaScan} captures via CUDA-event tracing and analyses efficiently using its multi-stage heuristic algorithm.

\subsubsection{Heterogeneous Computing for LLM Training}

As accelerator diversity widens, frameworks must seamlessly span GPUs, CPUs, NPUs, and emerging ISAs.  Jiang \emph{et al.} outline a unified architecture that shards model states across heterogeneous GPU/CPU clusters, leveraging smart replica placement and prioritized gradient staging to sustain bandwidth parity~\cite{Yan2024FlashFlex}.  NVIDIA's 2025 announcement of CUDA support for RISC-V CPUs signals further heterogeneity at the system-software layer, opening the door to bespoke edge chips orchestrating large-model workloads~\cite{Nvidia2025RiscV}.  Prior arts like FlashFlex~\cite{Yan2024FlashFlex, athlur2022varuna} solves training LLMs on heterogeneous clusters treat device diversity as a first-class scheduling signal, rather than forcing homogeneous parallelism across all participants.
\texttt{MegaFBD} builds on these insights by decoupling forward and backward passes, thereby freeing the scheduler to map lighter compute onto CPUs or other devices with poorer performance while reserving the fastest GPUs for compute-dense stages.

Collectively, these four research streams motivate the design choices in \texttt{MegatronApp}: trace-driven interpretability hooks, adaptive pipeline orchestration, fast straggler detection, and flexible device placement--features essential for pushing LLM training into the heterogeneous, trillion-parameter era.

%% file: MegaScan.tex
\section{\sd: Operator-Granular Tracing and Straggler Detection}

\subsection{Motivation and Goals}
As the parameter scale of Transformer-based large language models (LLMs) grows exponentially, model training has evolved from single-machine computation to a distributed, cross-node cooperative task. Training frameworks exemplified by \textit{Megatron-LM} make trillion-parameter model training possible through an organic combination of tensor parallelism, pipeline parallelism, and data parallelism.

Although these parallel strategies significantly boost training throughput, they also greatly increase system-level complexity; the training system becomes extremely sensitive to hardware stability and communication reliability as the number of nodes rises. In real-world deployments, transient faults (such as GPU down-clocking or link jitter) often trigger cascading delays that degrade training performance. More critically, because the communication topology is highly coupled, fault signals propagate along hidden paths with uncertain impact scopes, making root-cause localization particularly challenging.

Current mainstream performance-monitoring and anomaly-detection methods face three key challenges:
\begin{enumerate}
  \item \textbf{Lack of temporal-dependency modeling}. Traditional metrics (e.g., GPU utilization) only reflect local states and cannot capture causal chains between nodes. For instance, although NVIDIA's DCGM tool~\cite{nvidiaDCGM} can collect metrics such as memory usage and bandwidth, it struggles to tell influencers from those affected.
  \item \textbf{Missing communication-pattern modeling}. Most methods target general distributed systems and therefore fail to capture the highly structured communication semantics of training workloads. This results in high false-positive rates, especially under specific configurations such as tensor or pipeline parallelism.
  \item \textbf{Confounding of hardware and software anomalies}. Hardware issues-like GPU down-clocking or power fluctuations-exhibit temporal patterns similar to software factors such as data-distribution skew or batch-size changes, yet the detection strategies required for each are fundamentally different.
\end{enumerate}

\textbf{\sd} tackles these challenges by introducing a temporal-analysis and slow-node-detection framework at the granularity of application-level training operators. The system builds a lightweight tracing infrastructure centered on CUDA Events~\cite{nvidia_cuda_toolkit}, unifies visualization via the Chrome Tracing~\cite{chrome_tracing} format, and performs root-cause diagnosis through cross-rank dependency analysis, clock alignment, and bandwidth assessment. Compared with traditional approaches, \sd aligns more closely with actual training semantics while offering extremely low intrusiveness and strong scalability.

\subsection{Design}

\begin{figure}[ht]
  \centering
  \includegraphics[width=\linewidth]{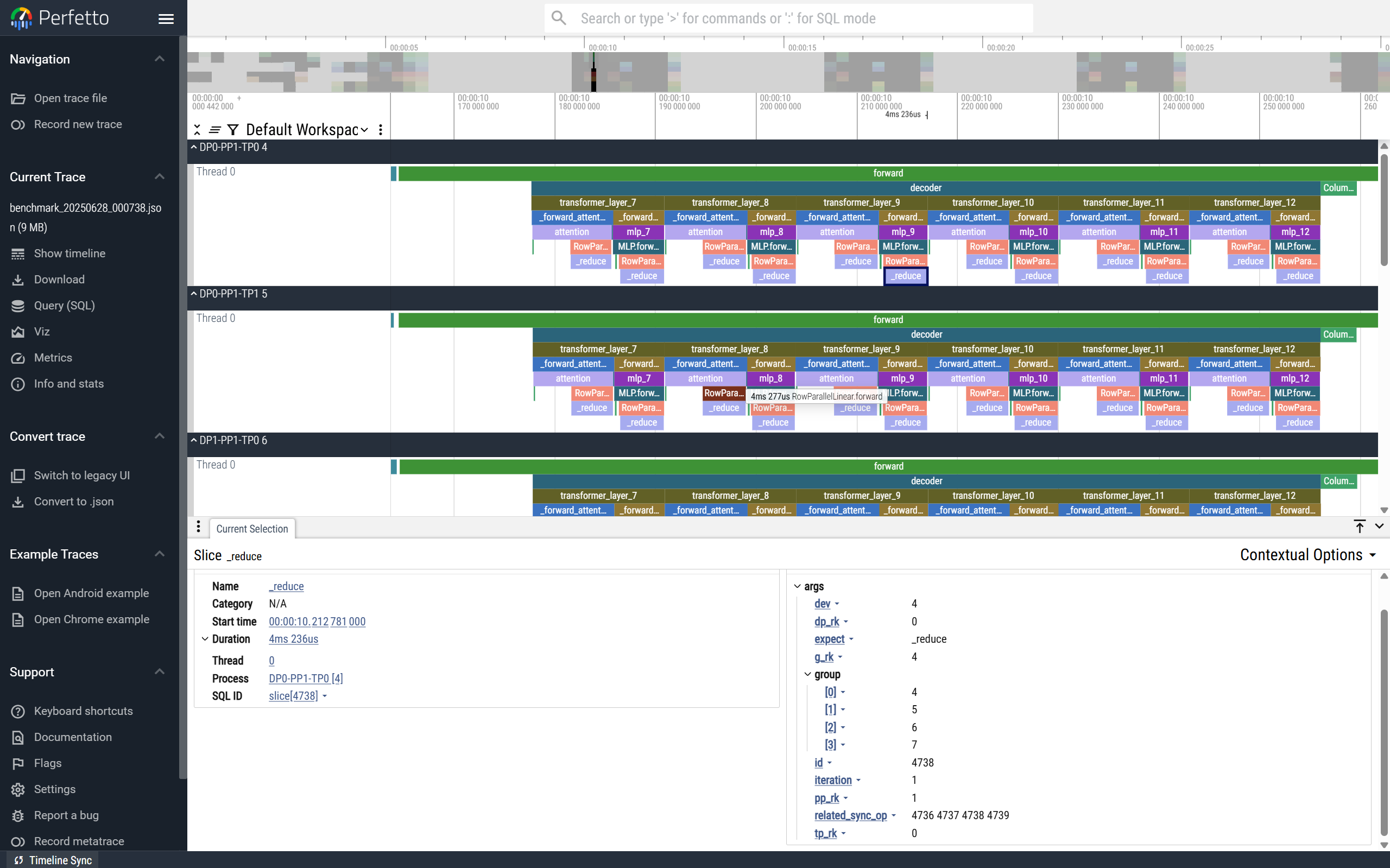}
  \caption{\sd visualizes the trace file using Chrome Tracing (or Perfetto UI).}
  \label{fig:trace}
\end{figure}

\para{Workload tracing.} \sd adopts a timing mechanism based on \emph{CUDA Events}.  
A CUDA event is a special marker, injected by the host into a CUDA stream, that is essentially an empty kernel.  
When the GPU's execution flow reaches this marker, it records the current timestamp.  
By inserting a \emph{start} event and an \emph{end} event in the same CUDA stream immediately before and after an operation under test (e.g.\ a compute kernel or a communication call), we can later query the elapsed time between the two events asynchronously via \texttt{cudaEventElapsedTime}.  
This time difference precisely reflects the true execution time of that operation on the GPU.  
Recording a CUDA event is lightweight, and fetching the timestamps as well as computing the difference can be done asynchronously, avoiding any synchronization overhead on the critical execution path.  
Consequently, the event-based approach captures the real latency of asynchronous operations with high fidelity while imposing negligible performance impact on the original training workload, making it an ideal foundation for the low-intrusion tracing framework required by our system.

Beyond recording event names and timestamps, the tracing framework allows additional \emph{metadata} (Arguments) to be attached to each event-such as the current micro-batch index, the amount of data involved in a communication, or the peer rank-by passing extra parameters to \texttt{tracers.scope}.  
These metadata are essential for later constructing a global dependency graph, understanding system behavior, and performing fault diagnosis.

\para{Log pre-processing.} During distributed training, each GPU process (\emph{rank}) independently produces its own trace, and rank 0 gathers all traces and persists them to disk. The gathering and persistence operations are conducted in a separate thread to alleviate training stalls.
After training, every rank has its own recorded event sequence as a JSON file.  
Because each rank timestamps events with its local GPU clock, directly merging these files does not yield an accurate global timeline.  
To obtain a consistent global view for cross-rank analysis, we aggregate and align the independent trace files.  
We implement an aggregation script that reads all per-rank JSON files and merges them-ordered by time-into a single JSON file conforming to the \emph{Chrome Tracing Format}.  
Originally designed for analyzing asynchronous events and performance bottlenecks inside the Chrome browser, this open JSON-based standard is now widely used for visualizing temporal event data.  
Users can load the merged trace in the built-in viewer (\texttt{chrome://tracing}) without additional software (as shown in Figure~\ref{fig:trace}).  
The viewer clearly displays, along a timeline, the start/end times, exact durations, concurrency, hierarchical nesting, and logical flow of events across different processes or threads (each rank is mapped to a separate process in our scenario).  
Users can zoom, pan, and search for specific events, examining the detailed parameters we carefully recorded-such as micro-batch IDs, communication volume, and rank lists-directly in the GUI.  
Moreover, thanks to its broad adoption, Chrome Tracing enjoys mature community support and toolchains for data sharing, conversion, and further development.

\para{Dependency reconstruction.} After obtaining the aggregated-yet not fully aligned-trace, a key step is to explicitly link synchronous communication operations executed on different ranks.  
In distributed training, synchronous collectives such as \texttt{AllReduce}, \texttt{AllGather}, and \texttt{Broadcast}, as well as point-to-point (\texttt{Send}/\texttt{Recv}) operations, require all participating ranks to reach a common synchronization point before the operation can complete.  
Identifying which concrete event instances form the \emph{same} communication operation is therefore crucial for subsequent analyses.  
To enable this, our tracer records, in addition to latency, the \emph{process-group} or \emph{peer-rank} information for every traced communication call.  
For collective operations we log the global ID list of all participating ranks; a single pass over the events then matches those that belong to the same communication instance.

\para{Timeline alignment.} Given the collections of events that logically constitute the same synchronous communication (stored in the \texttt{related\_sync\_op} attribute), we can exploit the fact that all participants in a synchronous call must \emph{logically} finish at the same moment before proceeding.  
This provides anchor points for aligning different ranks' timelines.  
We choose a reference rank (e.g.\ Rank~0) and iteratively align the other ranks to it, using the identified communication instances as calibration points.  
Because collectives occur frequently, the calibration points are dense, limiting clock-drift errors to the interval between two consecutive anchors and preventing long-term error accumulation.  
Under high-frequency communication patterns-such as in tensor parallelism or during intensive pipeline-stage exchanges-the alignment accuracy is correspondingly high.

\para{Anomaly analysis.} We devise a multi-stage heuristic algorithm to detect and localize down-clocked GPUs.  
The core insight is that the \emph{true} fault source should appear as the slowest member in \emph{every} synchronous group it joins, whereas ranks that merely suffer collateral slowdown will only lag because they are waiting for the faulty peer.

\begin{enumerate}
  \item \textbf{Cross-data-parallel comparison of peer operations}.  
        Data parallelism guarantees that ranks playing the same role-i.e.\ having identical PP and TP indices-execute identical sequences of compute kernels.  
        If a rank's execution time for a given kernel is significantly longer than that of its peer ranks, we mark that kernel instance as a \emph{slow operation}.  
        We then compute the proportion of slow operations per rank; ranks whose proportion exceeds a threshold become \emph{candidate} ranks.
  \item \textbf{Root-cause identification via collective synchronization}.  
        We further analyze candidates within their TP and DP collectives.  
        If a rank consistently \emph{starts} collective calls noticeably later than its peers-because its preceding computation is slower-and this pattern repeats across many collectives, it is likely the genuine fault source rather than a victim.
  \item \textbf{Root-cause identification via P2P latency}.  
        Pipeline parallelism complicates matters: the start times of \texttt{Send}/\texttt{Recv} between adjacent stages naturally differ due to pipeline scheduling and micro-batch asynchrony, even when the system is healthy.  
        Start-time comparison therefore cannot readily flag slow nodes.  
        Instead, we leverage the payload size (tensor size) and the observed latency recorded for each P2P transfer to compute the \emph{effective bandwidth}.  
        A true slow node often exhibits degraded bandwidth, either because its PCIe-NIC data path is impaired or because it prepares/consumes data slowly, especially during the \emph{warm-up} phase (1-forward-1-backward) where forward activations are sent downstream with few confounding interactions.
\end{enumerate}

\para{Fast data retrieval.} To efficiently query and analyze bandwidth data we use \emph{Perfetto SQL}~\cite{perfetto}.  
Perfetto, developed by Google, supports Chrome Tracing Format and offers a built-in SQLite-like query engine.  
We can import the aligned trace into Perfetto and write SQL queries to extract communication events, compute their effective bandwidths, and perform statistical analyses.

%% file: MegaFBD.tex
\section{\fbd: Heterogeneity-Aware Forward-Backward Decoupling}

\subsection{Motivation and Goals}

In the three-dimensional parallel paradigm adopted by Megatron-LM--tensor, pipeline, and data parallelism--the forward and backward phases are, by default, scheduled on the same GPU.  
Such binding ignores their structural differences in resource usage: the forward pass has more network transfers, whereas the backward pass involves gradient computation, activation recomputation, and optimizer updates.  
Consequently, the two phases exhibit highly dissimilar memory footprints, communication patterns, FLOP distributions, and power profiles.

Problems of co-locating forward and backward include: 
\begin{itemize}
  \item \textbf{Resource contention}: overlapped memory/bandwidth requests from the forward and backward passes can block the critical path.
  \item \textbf{Scheduling entanglement}: activations cannot be released early in the forward pass because the backward pass follows immediately.
  \item \textbf{Heterogeneous-resource under-utilization}: when CPUs, NPUs, or other devices are available, they cannot be flexibly reused.
\end{itemize}

To overcome these problems, we propose four core strategies in \fbd:
\begin{itemize}
  \item \textbf{Instance-level decoupled scheduling}: split the forward and backward phases into two logical processes, assign distinct ranks, and bind them to different resources to reduce coupling.
  \item \textbf{Heterogeneous resource mapping}: deploy the forward phase on devices with higher computation power.
  \item \textbf{Differential parallelism configuration}: assign a smaller parallel degree to the forward pass--leveraging lower GPU memory usage--to cut communication cost.
  \item \textbf{Thread-level coordination mechanism}: employ a communication coordinator to ensure necessary data synchronization between forward and backward, preventing deadlocks and redundant transfers.
\end{itemize}

Implemented as a low-intrusion enhancement atop vanilla Megatron-LM, \fbd adds a new dimension of resource scheduling, making it particularly suitable for heterogeneous clusters and resource-constrained environments.

\subsection{Design}

\begin{figure}[ht]
  \centering
  \includegraphics[width=0.8\linewidth]{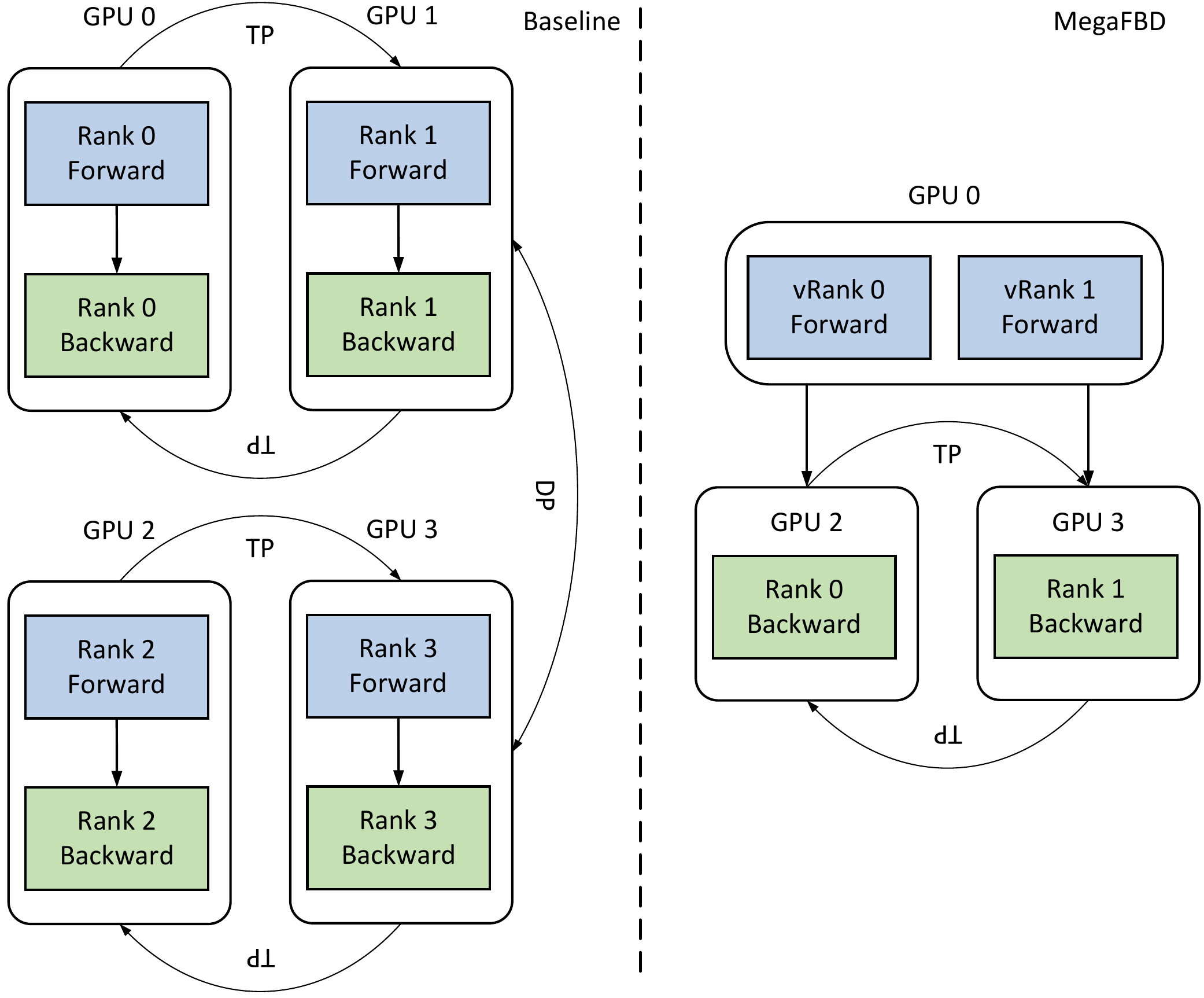}
  \caption{Forward/backward instance deployment of \fbd compared with that of existing training frameworks.}
  \label{fig:fbd}
\end{figure}

\para{Thread-level workers and virtual ranks.} To minimise changes to the original framework, \fbd keeps the fact that ``the forward and backward phases use different parallel configurations'' transparent to the training framework itself.  
Consequently, \fbd maintains \emph{two} parallel naming systems as shown in Figure~\ref{fig:fbd}:

\begin{itemize}
  \item \textbf{Virtual ranks}: every training thread owns a unique rank ID, and the forward and backward instances have the \emph{same} number of ranks.  
        Virtual ranks follow Megatron's original allocation rules, ensuring that model partitioning and computation proceed according to the framework's existing logic. 
  \item \textbf{Physical ranks}: the rank IDs actually held by each GPU.  
        Because the forward and backward phases may differ in their degrees of parallelism and compute capability, multiple training threads may reside on the \emph{same} GPU.  
        These intra-GPU threads need no heavy collective communication (they share the same device memory); any external communication is uniformly managed by the control thread of \fbd's \emph{communication coordinator}.
\end{itemize}

During initialization, physical ranks are created and assigned by the PyTorch launcher (process-level setup).  
Each process then spawns the required number of workers (thread-level setup), inheriting Megatron's rank-assignment policy.

\para{Communication coordinator.}
A common issue in multi-threaded collective communication is deadlock due to conflicting peer choices.  
Consider the scenario where Rank~0 wants to communicate with Rank~2, but Rank~2 intends to communicate with Rank~1.  
If we launch communications without verification, the control thread may start the Rank~0-2 transfer first; because Rank~1 shares the same control thread, its Rank~1-2 request is blocked, causing a deadlock.

The solution is to start a collective \emph{only} after confirming that all ranks in the group have issued the \emph{same} request.  
\fbd therefore introduces a \textbf{communication coordinator} that performs process-level synchronization and eliminates deadlocks.

When a worker thread needs to initiate a collective (e.g.\ all-reduce, broadcast, P2P), it first sends a request to its control thread.  
Control threads, which run permanently, exchange these requests within the relevant communication group.  
Once a control thread detects that \emph{all} participants have posted the same request, it launches the operation.

The coordinator handles only \emph{cross-GPU} communication.  
For each eligible communication group it maintains a bit-vector table of size $(\text{number of groups}) \times (\text{bit-vector length})$, where each bit vector maps one-to-one to all virtual ranks.  
The table is flattened into a 1-D tensor; different coordinators exchange this tensor via a global communication group, allowing them to determine how many members of each group are ready and to execute collectives in group order.  
The detailed steps are:

\begin{enumerate}
  \item \textbf{Registration}: every thread marks each planned collective by setting the bit corresponding to its virtual rank in the group's bit vector to~1.
  \item \textbf{Alignment}: threads periodically perform an \texttt{all-reduce} with bitwise~OR on the flattened vector.
  \item \textbf{Readiness check}: each thread compares the current bit vector with the group's \emph{expected} value (e.g.\ with eight threads, the group \{0,1,2,3\} expects \texttt{11110000}). If equal, the collective is ready.
  \item \textbf{Ordered execution}: threads execute the ready collectives in ascending group order, avoiding contention and starvation.
\end{enumerate}

This design exploits the three-dimensional parallel property of ``serial within groups, parallel across groups.''  
Bit-vector compression reduces scheduling overhead to $\mathcal{O}(G)$, where $G$ is the number of communication groups, and proves more efficient and scalable than traditional lock- or condition-variable-based thread synchronization schemes.

%% file: MegaDPP.tex
\section{\dpp: Dynamic Pipeline Scheduling}

\subsection{Motivation and Goals}

Pipeline parallelism is a key technique for boosting large-model training throughput: the model is partitioned into multiple sub-modules that execute concurrently on several GPUs in a pipeline fashion.  
The conventional \emph{1F1B} schedule--one forward pass and one backward pass per iteration--minimizes memory footprint but exposes three critical drawbacks:

\begin{enumerate}
  \item \textbf{Weight-update latency}: \texttt{AllReduce} for gradient synchronization can be issued only after \emph{all} backward stages finish, delaying the optimizer step.
  \item \textbf{Shrunken communication window}: forward outputs must be sent to the next stage immediately, leaving short overlapping windows between communication and subsequent computation.
  \item \textbf{Rigid scheduling}: the fixed order lacks adaptability to disturbances such as link-bandwidth fluctuations or GPU down-clocking.
\end{enumerate}

Prior studies have proposed various enhancements: \textit{PipeDream-2BW} employs double buffering to shorten pipeline flushes; \textit{ZB1P} compresses bubbles via an optimized schedule graph; \textit{BitPipe} introduces bidirectional interleaving to raise throughput.  
However, all of them retain the 1F1B assumption, keeping the schedule rigid and ill-suited for dynamic environments.

\dpp tackles this challenge with an \textbf{elastic, adaptive pipeline-scheduling framework} that supports dynamic traversal orders and resource-aware decisions within the Megatron-LM ecosystem.  Its key designs are:

\begin{itemize}
  \item \textbf{Traversal-policy switching}: flexibly alternate between \emph{depth-first} (advancing the same data through multiple model blocks) and \emph{breadth-first} (advancing multiple data items through the same block) strategies.
  \item \textbf{Resource-aware scheduling}: prioritize tasks based on real-time memory usage and network-bandwidth conditions, hiding communication bottlenecks at low overhead.
  \item \textbf{Asynchronous-communication support}: a lightweight parallel scheduler plus an async P2P communication library fuse computation and transfer to maximize device utilization.
\end{itemize}

\dpp delivers a decisive breakthrough over traditional static schedules, granting training systems greater robustness and scalability under complex parallel settings.

\subsection{Design}

\begin{figure}[ht]
  \centering
  \includegraphics[width=0.8\linewidth]{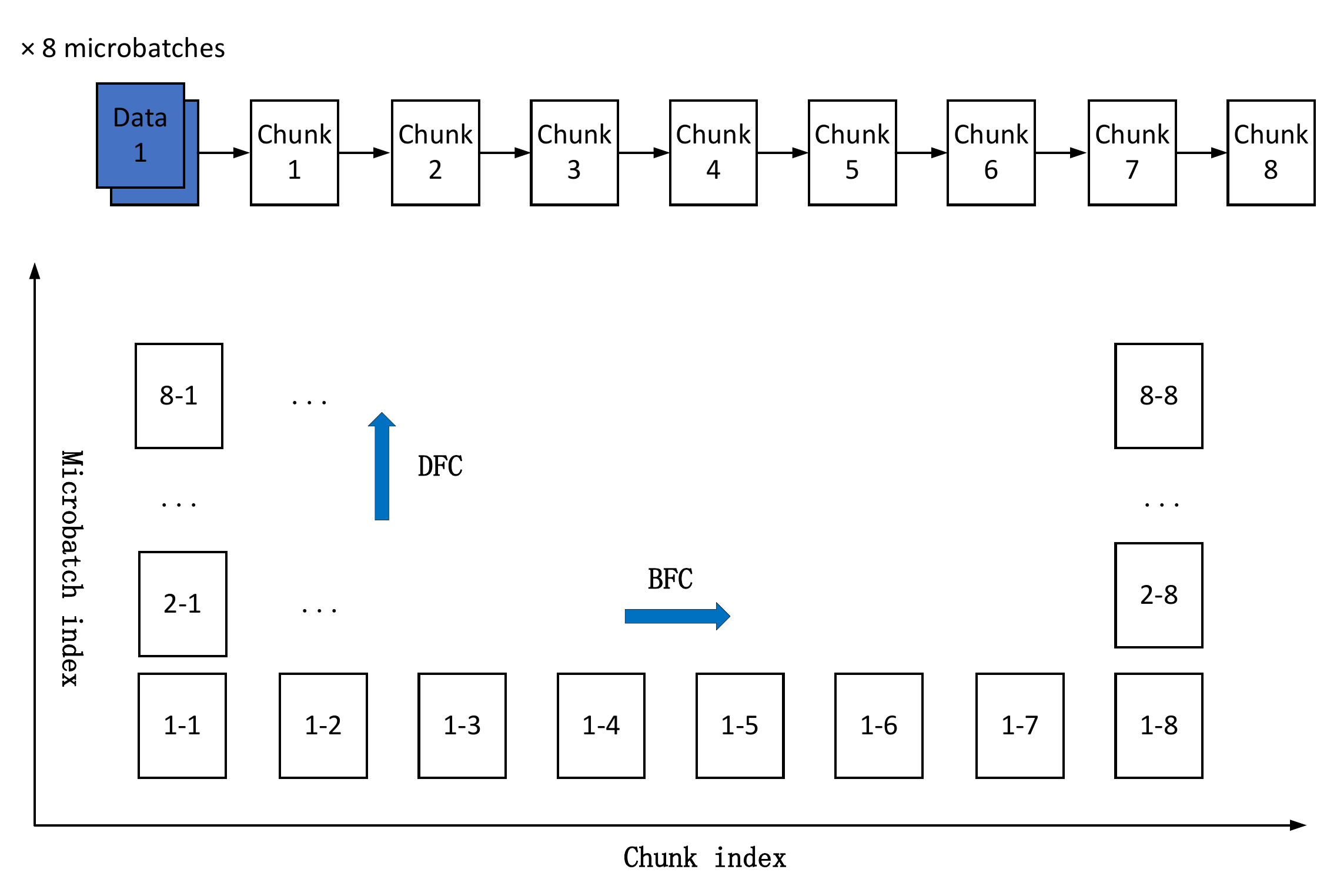}
  \caption{Training models with 8 chunks on 8 microbatches of data require 64 computation tasks per iteration. $i$-$j$ denotes the task that processes the $i$-th microbatch on chunk $j$. \dpp scheduler arranges these tasks into an $8 \times 8$ matrix and executes them successively along either the micro-batch (row) or model-chunk (column) dimensions.}
  \label{fig:dpp}
\end{figure}

\para{Parallel scheduler.} In pipeline parallelism each rank runs an independent \emph{parallel scheduler} that decides the execution order of all compute and communication steps.  
If we view every compute task of one training iteration as a two-dimensional matrix indexed by \((\text{model\_chunk\_id}, \text{microbatch\_id})\)--that is, producing an output on a given model shard for a given data shard--then a schedule is simply a \emph{traversal order} over this matrix.  
\textbf{\dpp} supports two traversal strategies:

\begin{itemize}
  \item \textbf{Depth-First Computation (DFC)}: process the \emph{same} micro-batch across \emph{different} model chunks first.  
        This lets the corresponding backward pass start earlier, releases activation memory sooner, and lowers the GPU-memory peak.
  \item \textbf{Breadth-First Computation (BFC)}: process \emph{different} micro-batches on the \emph{same} model chunk first.  
        Portions of a chunk thus finish all their computation sooner, enabling earlier gradient synchronisation, while significantly delaying downstream consumption of activation outputs--hence reducing network-communication pressure.
\end{itemize}

\dpp allows users to choose either scheme explicitly or to adopt BFC in a \textit{best-effort} fashion as long as it does not cause out-of-memory (OOM) errors (Figure~\ref{fig:dpp}).

\para{Lightweight communication library.} We implement an intra-/inter-node P2P library atop shared memory and RDMA interfaces.  
Unlike mainstream NCCL, our API permits a single device to issue \emph{concurrent, asynchronous} P2P transfers, so a compute thread can always pick the highest-priority ready input for execution.  
Concretely:

\begin{itemize}
  \item \textbf{Four buffers} store, respectively, (i) forward tensors received from the network, (ii) forward outputs to be sent, (iii) backward tensors received, and (iv) backward outputs.
  \item \textbf{Two task queues} correspond to the sender and receiver.  
        When the main thread invokes a communication call, the task is enqueued.
  \item \textbf{Dedicated worker threads} dequeue tasks, perform the actual send or receive, and copy tensors to the designated addresses.
  \item The \textbf{main thread} tracks completion of all queued tasks, thus always knows which concurrent transfers have finished.
\end{itemize}

For collective communications such as \texttt{all-reduce}, \texttt{all-gather}, and \texttt{reduce-scatter}, we continue to rely on NCCL.

%% file: MegaScope.tex
\section{\vs: Real-Time Interactive Visualization for LLM Interpretability}

\subsection{Motivation and Goals}
Interpretability and training-process analysis for large language models are becoming research hot-spots.  
Yet once parameter counts exceed the hundreds-of-billions or even trillions, the training state space becomes enormous and highly dynamic, posing unprecedented challenges for visualization systems.  
On the one hand, researchers struggle to extract meaningful behaviours and patterns from surface-level metrics and to
uncover the causal chain between parameter updates and performance
evolution.   
On the other hand, directly logging and displaying internal states--activations, attention matrices, gradients--can incur significant I/O and communication overhead, which may severely harm training efficiency in distributed settings.

Mainstream tools reveal several limitations when applied to LLM
visualization:

\begin{itemize}
  \item \textbf{Fixed metric dimensions}: tools such as \textit{TensorBoard} display only framework-defined metrics and hardly support user-defined training signals.
  \item \textbf{Tight coupling and manual export}: solutions like \textit{BertViz} require users to insert extra code to dump tensors, breaking the training code structure; for large models the export cost is prohibitive.
  \item \textbf{Lack of interaction and injection}: most tools provide only static or streaming views, without the ability to inject intervention signals during training--hence no support for behavioural debugging or interpretability experiments.
\end{itemize}

Key features of \vs include:

\begin{itemize}
  \item \textbf{Dynamic sampling and asynchronous processing}: users
        define observation points (layer / token / metric) via a registration API.  The system captures the requested intermediate tensors on demand and caches them asynchronously, minimizing interference with the training path.
  \item \textbf{Hierarchical compression and smart caching}: activations and attention tensors are compressed online on the host side using aggregate statistics (e.g.\ max, mean, sparsity).  
        Multi-level caching strategies control bandwidth and storage pressure.
  \item \textbf{Explorable interpretability UI}: the front-end presents multi-level, multi-granularity views--including attention heatmaps, representation trajectories, and token evolution--enabling full-chain insight from token behaviour to model mechanisms.
  \item \textbf{Perturbation injection and controlled experiments}: users can inject interventions at specific layers--such as replacing an attention matrix or adding perturbations--for rapid experiment design, enhancing research flexibility.
\end{itemize}

\vs greatly improves the observability of the training process and offers a powerful experimental tool for LLM architecture studies, robustness testing, and behavioural analysis.

\subsection{Design}

\begin{figure}[ht]
  \centering
  \includegraphics[width=\linewidth]{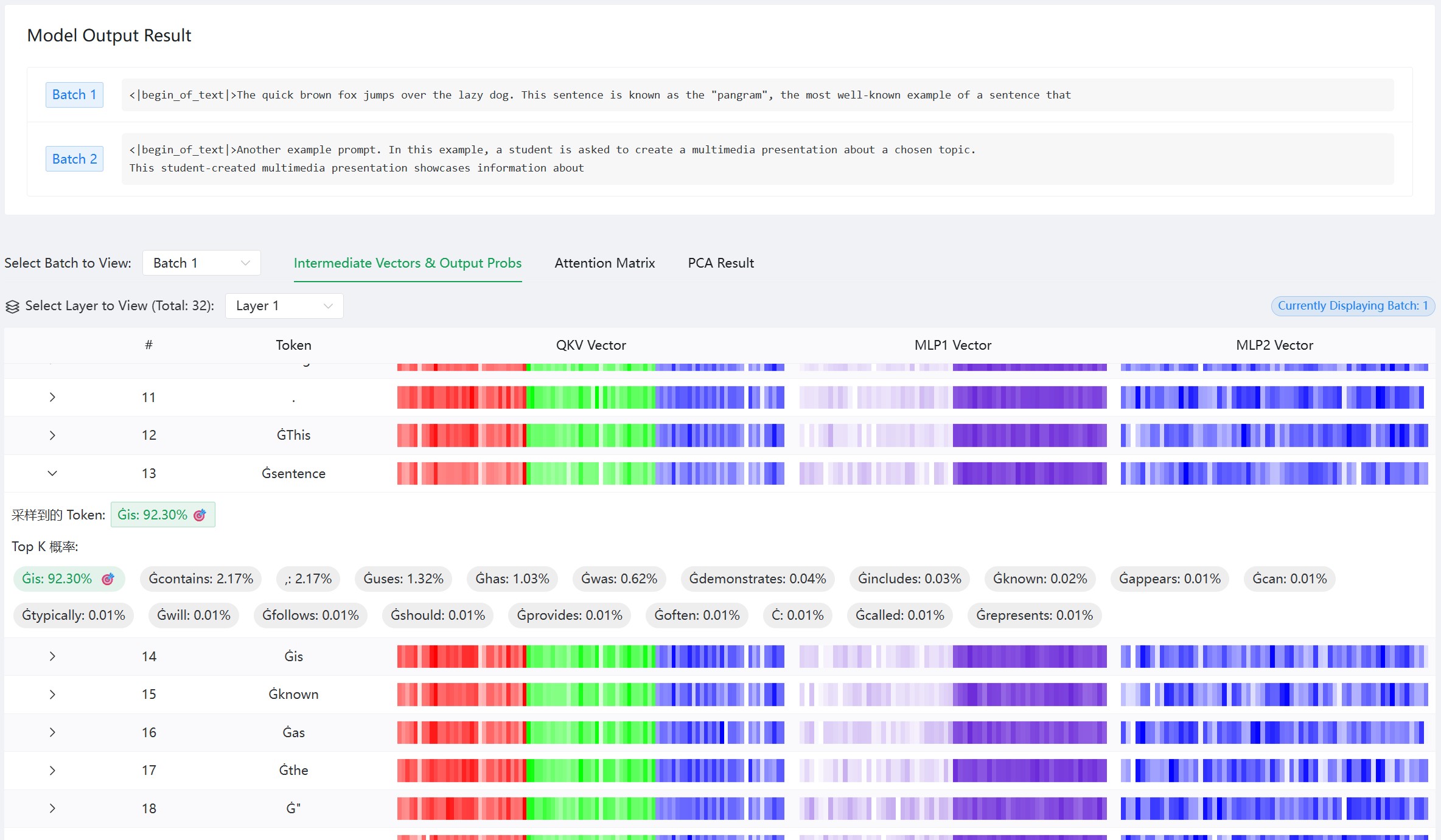}
  \caption{An example of visualizing text generation. \vs displays the visualization results token-by-token. In the first tab, the intermediate vector heatmaps are displayed and the output probabilities are shown in the expandable sections.}
  \label{fig:visual}
\end{figure}

\begin{figure}[ht]
  \centering
  \includegraphics[width=\linewidth]{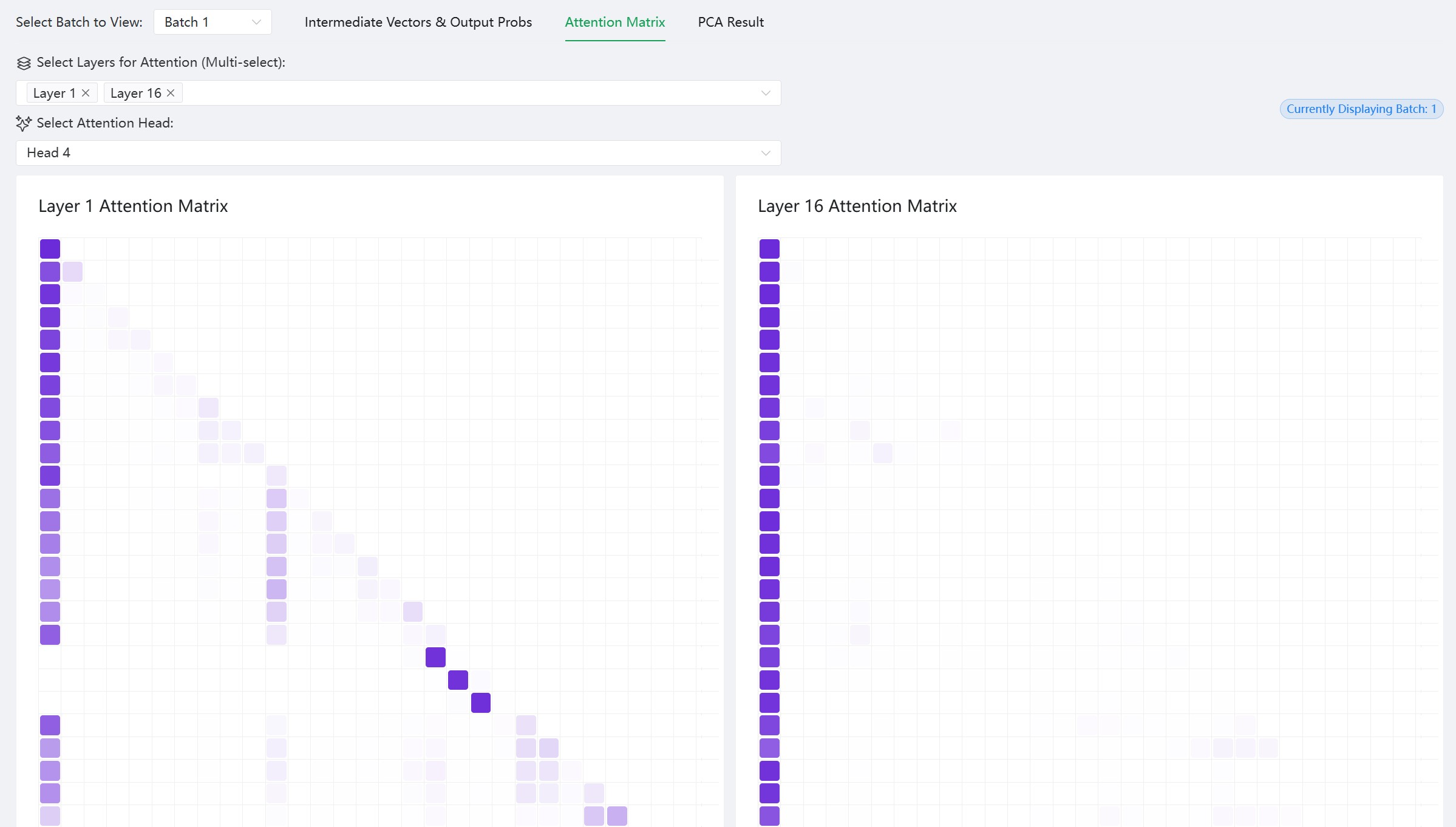}
  \caption{An example of visualizing attention scores.  Users can select the layer and attention head they wish to inspect.}
  \label{fig:attention}
\end{figure}

\begin{figure}[ht]
  \centering
  \includegraphics[width=\linewidth]{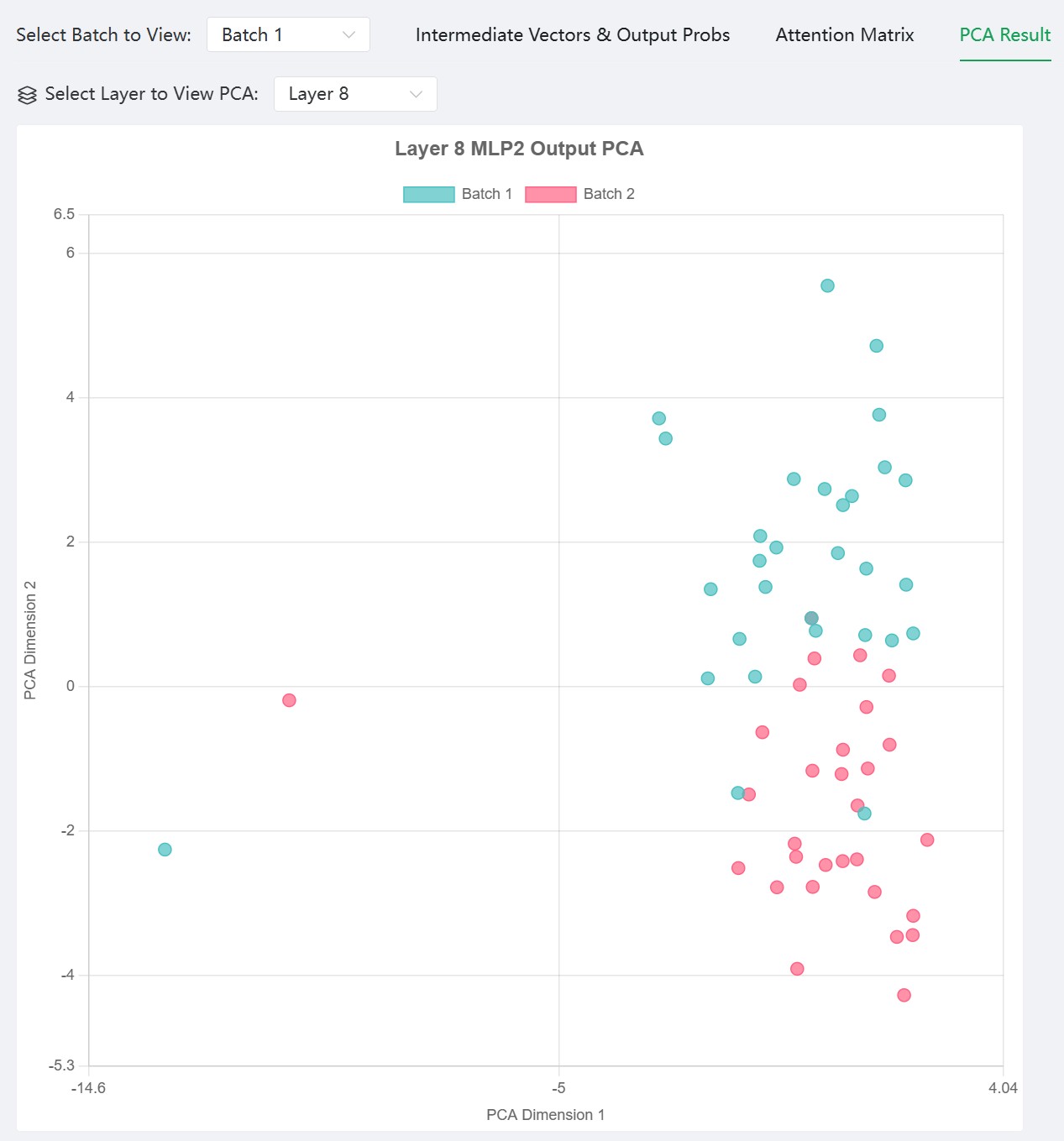}
  \caption{An example of visualizing the PCA dimensionality reduction feature.Users can visually inspect the clustering of tokens and understand how the model groups similar concepts. The displayed layer can also be selected.}
  \label{fig:pca}
\end{figure}

\para{Real-time generation and synchronized visualization.}
After the user enters any prompt, the front-end displays the model's decoding process in a \emph{token-by-token refresh} mode while simultaneously presenting the back-end's captured internal states (activations, attention maps, gradients, \emph{etc.}), achieving a one-to-one correspondence between observed behaviour and underlying mechanisms. Figure~\ref{fig:visual} displays a visualization example of \vs.

\para{In-Depth analysis of intermediate results.}
To highlight potential distribution drifts or outliers, \vs renders density heatmaps for crucial intermediate tensors such as the query, key, and value vectors \(\mathbf{Q},\mathbf{K},\mathbf{V}\), the outputs of the MLP sublayers, and the residual branches.  For attention diagnostics, the system retrieves weights at both layer and head granularity, then displays their temporal evolution through interactive heatmaps and short animations; researchers can thus juxtapose attention patterns arising from different prompts (Figure~\ref{fig:attention}).  During every ``predict-next-token'' step, the interface shows the chosen token alongside its probability and overlays a top-\(k\) bar chart that makes the entire decision distribution visually explicit.

\para{Interactive exploration tools.}
A quick-access panel lets users pivot fluidly across layer, head, token, and batch dimensions; these selections are tied to a timeline component so that the decoding process can be replayed or scrubbed freely.  High-dimensional hidden states are also projected onto two dimensions via PCA (Figure~\ref{fig:pca}), and an interactive scatter plot traces each token's trajectory, exposing geometric relationships between tokens and their originating prompts.

\para{Pluggable perturbation-injection framework.}
Before critical parameter tensors are written back to memory, \vs can inject random bit flips or Gaussian noise, enabling systematic studies of how storage faults undermine model robustness.  During the forward pass, researchers may add noise fields or masking functions to the output of any chosen layer, making it straightforward to explore the behavioural impact of reduced numerical precision.  At the system level, a constant offset can be introduced into the results transferred between layers, which emulates cross-device quantization errors or persistent link jitter and helps quantify the model's resilience to communication anomalies.

%% file: Evaluation.tex

%% file: Conclusion.tex
\section{Conclusion}

In this work, we propose \textbf{\sysname}, a cohesive, production-ready toolchain that augments the training, debugging, and visualization capabilities of Megatron-LM at trillion-parameter scale. By integrating four purpose-built modules--\textbf{\sd}, \textbf{\fbd}, \textbf{\dpp}, and \textbf{\vs}--the framework transforms large-model development from a monolithic, opaque process into an analyzable, tunable, and highly efficient workflow.

\begin{enumerate}
    \item \textbf{\sd}: Fine-grained performance tracing and slow-node detection with near-zero overhead, enabling practitioners to isolate system bottlenecks in minutes rather than hours.
    \item \textbf{\fbd}: Decouples forward and backward passes to unlock heterogeneous resource utilization, reducing peak GPU memory and increasing overall throughput in mixed-CPU/GPU clusters.
    \item \textbf{\dpp}: Introduces adaptive pipeline scheduling that reacts to runtime imbalances, flattening straggler effects and improving hardware utilization across deep pipeline stages.
    \item \textbf{\vs}: Provides customizable metrics collection and real-time visualization, turning billions of training events into actionable insights through an intuitive dashboard.
\end{enumerate}

Together, these components establish a new operational baseline for large-scale language-model training:

\begin{itemize}
    \item \textbf{Higher efficiency}: Double-digit gains in throughput and cluster utilization.
    \item \textbf{Deeper observability}: Unified, operator-level visibility into computation and communication behavior, together with real-time and customized metrics collection on intermediate results.
    \item \textbf{Rapid diagnosis}: Automated anomaly detection and root-cause analysis that shorten mean time to recovery.
    \item \textbf{Future-proof extensibility}: A modular, open-source architecture ready for emerging hardware and novel parallelism strategies.
\end{itemize}

Looking forward, we plan to extend \sysname with:

\begin{itemize}
    \item Native support for fat failover after anomaly detection.
    \item New features and optimizations inherited from Megatron-LM.
    \item Additional tracing and diagnosis support for inference scenarios.
\end{itemize}

By lowering the barrier to efficient, transparent, and fault-tolerant training, \textbf{\sysname} positions itself as an indispensable companion for researchers and engineers pushing the frontier of large language models.